\documentclass[usenatbib]{mn2e}

\usepackage{rotating}
\usepackage{color}
\usepackage{pdflscape}
\usepackage{graphicx}
\usepackage{txfonts}

\newcommand{\aap}{A\&A} 
 
\newcommand{\mnras}{MNRAS} 
\newcommand{\apj}{ApJ} 
\newcommand{\araa}{ARA\&A} 
\newcommand{\apjl}{ApJ}
\newcommand{\apjs}{ApJS}
\newcommand{\aj}{AJ}
\newcommand{\pasp}{PASP}

\newcommand{\teff}{$T_{\mbox{\scriptsize eff}}$}
\newcommand{\kms}{$\mbox{kms}^{-1}$}
\newcommand{\logg}{log\,$g$}

\newcommand{\msun}{M_\odot}
\newcommand{\vrms}{v_{\rm rms}}
\newcommand{\trh}{\tau_{\rm rh}}
\newcommand{\rh}{r_{\rm h}}

\title[]{Search for Associations Containing Young stars (SACY): Chemical tagging IC~2391 
\& the Argus association\thanks{Based on observations obtained at the European Southern Observatory, Paranal, Chile 
(ESO program 082.C-0218 and on observations made under the ON-ESO agreement for the joint operation of the 1.52\,m ESO telescope 
and at the  Observat\'{o}rio do Pico dos Dias operated by MCT/Laborat\'{o}rio 
Nacional de Astrof\'{\i}sica (LNA/MCT),  Brazil.}} 

\author[]{G. M. De Silva$^{1}$\thanks{E-mail: gayandhi.desilva@aao.gov.au}, V. D'Orazi$^{2,6}$, C. Melo$^{3}$, 
C. A. O. Torres$^{4}$, M. Gieles$^{5,7}$, G. R. Quast$^{4}$, \and M. Sterzik$^{3}$\\\\
$^{1}$Australian Astronomical Observatory, 105 Delhi Rd, NSW 2113, Australia\\
$^{2}$Macquarie University Research Centre in Astronomy, Astrophysics \& Astrophotonics NSW 2109, Australia; \\
Department of Physics \& Astronomy, Macquarie University, NSW 2109, Australia\\
$^{3}$European Southern Observatory, Casilla 19001, Santiago 19, Chile\\
$^{4}$Laborat\'{o}rio Nacional de Astrof\'{\i}sica/MCT, 37504-364, Itajub\'{a}, Brazil\\
$^{5}$Institute of Astronomy, University of Cambridge, Madingley Road, Cambridge CB3 0HA, UK\\
$^{6}$Monash Centre for Astrophysics, School of Mathematical Sciences, Building 28, Monash University, VIC 3800, Australia\\   
$^{7}$Department of Physics, University of Surrey, Guildford, GU2 7XH, UK
}

\begin{document}



\maketitle

\label{firstpage}

\begin{abstract}
We explore the possible connection between the open cluster IC 2391 and the unbound Argus association identified by the SACY survey. 
In addition to common kinematics and ages between these two systems, here we explore their chemical abundance patterns to confirm 
if the two substructures shared a common origin. 
We carry out a homogenous high-resolution elemental abundance study of eight confirmed members of IC~2391 as well as six members 
of the Argus association using UVES spectra. 
We derive spectroscopic stellar parameters and abundances for Fe, Na, Mg, Al, Si, Ca, Ti, Cr, Ni and Ba. 

All stars in the open cluster and Argus association were found to share similar abundances with the scatter well within the uncertainties, 
where [Fe/H] = $-$0.04 $\pm$ 0.03 for cluster stars and [Fe/H] = $-$0.06 $\pm$ 0.05 for Argus stars. 
Effects of over-ionisation/excitation were seen for stars cooler than roughly 5200~K as previously noted in the literature. 
Also, enhanced Ba abundances of around 0.6 dex were observed in both systems. 
The common ages, kinematics and chemical abundances strongly support 
that the Argus association stars originated from the open cluster IC 2391. 
Simple modeling of this system find this dissolution to be consistent with two-body interactions.
\end{abstract}
 

\section[]{Introduction}
\label{section:introduction}

The vast majority of stars that we observe in any given galaxy are field
stars, a complex mixture of stellar populations of different ages
and metallicities. The present view is that star formation taking place
in molecular clouds gives rise to groups of stars - stellar associations and open clusters.
These groups of stars must dissolve with time to connect star formation with the field population.\\

Based on near-infrared studies of embedded clusters, \citet{lada03} suggest that less than about 10\% of the clusters formed in molecular clouds survive longer than 10 Myr. They propose that this ``infant mortality" of clusters is due to feedback of massive stars that expels the gas from the star forming region leaving the stars behind with super-virial velocities. For compact systems of a few hundred stars the dynamical time-scale is only a few Myrs. This means that cluster expansion and star loss by two-body relaxation is also a relevant process at these ages for such systems \citep{moeckel2012,gieles2012}. On longer time-scales (few 100 Myrs) various processes such as the Galactic tidal field and interactions with Giant Molecular Clouds (GMCs) become important. The paucity of open clusters in the solar neighbourhood with ages between 0.5 and a few Gyrs was already noted in the late 50Õs \citep{oort58}. \citet{spitzer58} analytically derived the cluster disruption times due to encounters with GMCs, showing that such phenomena would account for the scarcity of Galactic clusters older than $10^9$ years. A more detailed treatment of the problem taking into account other disruptive agents has been carried out in the literature \citep[e.g.][]{lamers06}.

\cite{torres06} reported the results of a high-resolution optical spectroscopic survey to search for associations
containing young stars (SACY) among optical counterparts
of ROSAT All-Sky X-ray sources in the southern hemisphere.
Using the method described in \cite{torres06}, \cite{torres08} identified nine new young associations,
namely, Chamaleontis (ChA), TWHydrae (TWA), $\beta$ Pictoris,
Octants (OctA), Tucana-Horologium (THA), Columba (ColA),
Carina (CarA), Argus (ArgA), and AB Doradus (ABDA). The age span derived based on the
isochrone fitting and the Li abundance analysis ranges from 6-70~Myr \citep{dasilva09}.\\

Of relevance to the context of cluster disruption discussed above, are 
two young associations identified in the SACY survey which appear to be associated with known open clusters. 
These are the Argus association and the open cluster IC~2391 and the $\epsilon$ Cha Association and the
$\eta$~Cha cluster \citep{mamajek00,torres08}. 
In this paper we investigate in detail the possible connection between the Argus Association and the young open cluster IC~2391. \\

In particular, we can test for a common origin of the association and cluster by examining the stellar elemental abundances. 
If the abundance analysis reveals high chemical homogeneity among the association stars  comparable to the open cluster abundance patterns, 
it supports a common origin scenario where the Argus association stars were once bound members of open cluster IC~2391. 
This assumes that the proto-cluster gas cloud was sufficiently mixed hence all stars born from the same site should share 
the same elemental abundance pattern. 
This concept of elemental abundance matching is often referred to as {\it chemical tagging} \citep{fbh}. 
Conversely if the abundances of the Argus association stars are found to be either heterogeneous, 
matching the general disk stellar population (e.g. the Hercules stream, see \cite{hercules}) or they do 
not match the abundance patterns of IC~2391, then it is more likely that these are two separate systems even if they share the same kinematics.\\

Given that disk dynamics make the kinematical information somewhat unreliable to explore the history of these stellar structures, 
for the first time we carry out a high resolution elemental abundance analysis of the Argus association in a homogeneous manner 
with stars in IC~2391 in order to carry out chemical tagging. 
This will also be the first attempt at chemical tagging a young stellar system. 
All published examples of chemical tagging in the disk are for stellar systems older than the Hyades \citep{wolf630, hr1614}.\\

In Section \ref{sect:2}, we briefly describe the SACY survey and present space velocities and spatial locations of the stars in the two systems. 
In Section \ref{sect:3} we present a detailed elemental abundance study of stars in both the Argus association and open cluster IC~2391. 
In Section \ref{sec:disc} we discuss our results in terms of chemical tagging the two stellar systems. 
We conclude with a discussion of possible disruption mechanisms given the evidence gathered in the previous sections.




 \begin{figure*}
\begin{center}
\begin{tabular}{c}
\resizebox{1.0\hsize}{!}{
\includegraphics[draft=False]{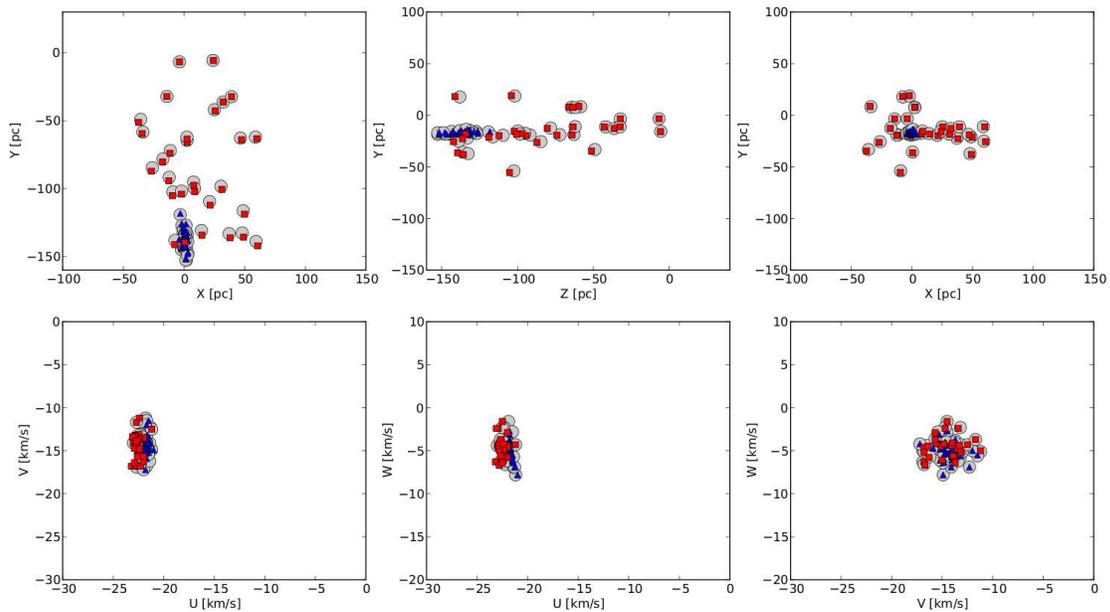}}
  \end{tabular}
    \end{center}
\caption{Combinations of the UVWXYZ--space derived based on the convergence method for the Argus Association stars (red squares), the open cluster IC~2391 (blue triangles) and when the two systems are considered as a single association (gray circles). Well defined clustering can be seen in both kinematical and spatial coordinates, with no significant difference when considering the two stellar system independently or together.} 
\label{fig:arxyz}
\end{figure*}

\section{Identifying the Argus association in the SACY survey}\label{sect:2}

\subsection{The convergence method}

Usually we think about an association as being a group of stars appearing concentrated in a small
volume in space sharing some common properties such as age, chemical composition, distance and kinematics. 
However, if such a group is close enough to the Sun, its
members will appear to cover a large extent in the sky (as an example,
Orion at 50~pc would cover almost the whole sky).  Thus, to find a
group, projected spatial concentrations (i.e., in terms of right
ascension and declination only) and proper motions may not be enough.
A better criterion is to look for objects sharing similar heliocentric
space motions (UVW) all around the sky (where we use U positive towards the
Galactic center, V positive in the direction of Galactic rotation, and W positive in the direction of the North Galactic Pole).\\

\citet{torres06} describe in some detail the convergence method developed
to search for members of an association. 
This method examines the stars in the hexa-dimensional space,
UVWXYZ, as defined by the space motions relative to the Sun  and the physical space coordinates centered
on the Sun (XYZ, in the same directions as UVW). We represent with m$_v$, M$_v$ and M$_{v,iso}$ the apparent visual magnitude,
the resultant absolute magnitude with the distance obtained from the convergence
method,  and the absolute magnitude given by  the adopted isochrone for the $(V-I)_C$ stellar color;
and   $\mu_\alpha$, $\mu_\delta$ and V$_r$ are the proper motions and the radial velocity.
In brief, if there is no reliable trigonometric distance available\footnote{We consider the trigonometric
parallaxes as unreliable if they have errors larger than 2~mas.},
the convergence method finds the distance (d) for each star in the sample
that minimizes the F value of Equation~\ref{eq:f}. The first term is a photometric distance modulus and
the second is a kinematical one. The method needs, as input, an assumed age and initial velocity values
$(U_0, V_0, W_0)$ for the proposed association, and a cutoff value for F above which stars should be considered spurious. 
This cutoff value varies for each association but usually we begin with 3.5
(this approximately means 0.7 magnitudes for the distance modulus and 3~km~s$^{-1}$ for the velocity modulus).\\

The method is iterative, and for each iteration a list of stars with new  $(U_0, V_0, W_0)$ is obtained.
The process ends when the list of stars and the velocities $(U_0, V_0, W_0)$ do
not change significantly.

\begin{eqnarray}
\label{eq:f}
\nonumber \lefteqn{F(m_v,\mu_\alpha,\mu_\delta,V_r;d) =}\\
&[p\times(M_v-M_{v,iso})^2+(U-U_0)^2+(V-V_0)^2+(W-W_0)^2]^{1/2}
\end{eqnarray}
where $p$ is a constant weighting the importance of the evolutionary
distance with respect to the kinematic distance.
Usually, if there are several stars with trigonometric 
parallaxes among the candidates, we use $p$=0. But if there are no stars with reliable parallaxes, 
as is the case for the IC~2391, the method may not converge, so we used $p$=20 for this case. \\

We also define a membership probability to verify the compactness of the association  
in 7 dimensions, namely UVWXYZ and $(M_v-M_{v,iso})$. 
Normalized values of these 
variables are computed by subtracting their mean values and dividing by their dispersions. 
A quantity K is obtained as the quadratic sum of these dimensionless variables and the 
probability is computed as 
\begin{eqnarray}
\label{eq:p}
\lefteqn{P=1-(erf(k))^7} 
\end{eqnarray}
where $erf$ is the error function. 

Finally, we also take into account the Li content of the candidate stars to see if it is 
compatible with the Li depletion for its age \citep{neu97}. 

\begin{landscape}
\begin{table}
{
\caption{Revised list of high probability members for the Argus association$^{1}$}  
\label{table:argus}
\begin{tabular}{lccrrrrlrrlrrr@{\hspace{1.5pt}}r@{\hspace{1.5pt}}rr@{\hspace{1.5pt}}r@{\hspace{1.5pt}}rrl}


\hline

Name  &  $\alpha$&$\delta$&$\mu_\alpha$&$\mu_\delta$&$~~~V_r$&Vsini&P$_{phot}$&$V_J$ & V-I$_C$ &SpT &E$_{Li}$& $\pi_{kin}$&\multicolumn{3}{c}{$(U,V,W)$}&\multicolumn{3}{c}{$(X,Y,Z)$}&Prob&notes\\
     &        2000&2000 &\multicolumn{2}{c}{~~mas yr$^{-1}$}&\multicolumn{2}{c}{~km s$^{-1}$}&days&	 & &    &m\AA & mas     &\multicolumn{3}{c}{km s$^{-1}$}&\multicolumn{3}{c}{pc}     &     &	 \\

\hline\hline
(*)CD-29 2360	&05 34 59.2	&-29 54 04	&17.1	&33.1 	&26.0     	        &&{\it3.519}&{\it10.53} &{\it1.11}	&K3Ve           &180          &14.8   &-22.7 & -16.6 & -5.2   &-34.9 &-47.9 & -32.4	&0.72 &K\\
(*)AP Col   	&06 04 52.2	&-34 33 36	&{\it27.3}&{\it340.9}&{\it22.4}	     &{\it11~~~}   &&{\it12.96} &{\it3.36}	&{\it M4Ve}	&{\it280}     &{\it119.2}&-22.0&-13.6&-4.4	&-3.7  & -6.7 & -3.4 	&0.90 & 1 \\
CD-56 1438 	&06 11 53.0	&-56 19 05	&-2.5	&38.3 	&14~~~               &130~~~&{\it0.418848}&{\it11.32} &{\it0.93}	&K0V  &230          &8.7 	&-21.9 & -11.2 & -5.1	&-9.1  &-101.3& -53.3	&0.68 &K \\
CD-28 3434	&06 49 45.4	&-28 59 17	&-11.3  &20.8 	&26.8             	&6.4&{\it3.823} &{\it10.62}  &{\it0.85}	&G7V      &230          &9.8 	&-23.1 & -16.3 & -7.0	&-51.2 &-85.3 & -23.1	&0.45 &K  \\
CD-42 2906	&07 01 53.4	&-42 27 56	&-13.8  &35.5 	&23.6              &10.8&{\it3.96 } &{\it10.64} &{\it1.02}	&K1V       	&275        &10.7	&-22.8 & -16.8 & -6.3	&-26.5 &-85.8 & -26.0	&0.75 &K  \\
CD-48 2972	&07 28 22.0	&-49 08 38	&-26.7  &44.4 	&21.1              &52.0&{\it1.0373}&{\it 9.85}  &{\it0.86}&G8V       	&250        &12.6	&-22.3 & -16.6 & -7.3	&-11.9 &-76.0 & -20.0	&0.88 &K  \\
(*)HD 61005  	&07 35 47.5	&-32 12 14	&{\it-55.7}&{\it74.6}&{\it22.5}	&{\it8.2}&{\it5.04}&{\it8.22 } &{\it0.80}	&{\it G8V }	&{\it171}     &{\it28.3}&-23.0&-14.1&  -4.4   &-14.1 &-32.2 & -3.5&0.73&H,2 \\
CD-48 3199	&07 47 26.0	&-49 02 51	&-22.4  &36.1 	&18.5              &24.7&{\it2.192 }&{\it10.56}  &{\it0.86}&G7V       	&230        &10.0	&-22.2 & -15.1 & -5.0	&-12.9 &-96.7 & -20.2	&1.00 &K  \\
CD-43 3604	&07 48 49.8	&-43 27 06	&-26.8  &39.6 	&21.4              &40.4&{\it0.4706}&{\it11.13}  &{\it1.40}&K4Ve      	&320        &12.2	&-22.6 & -16.4 & -4.8	&-17.5 &-79.1 & -12.6	&0.82 &K  \\
TYC 8561-0970-1&07 53 55.5	&-57 10 07	&-18.3  &26.9 	&15.8              &5.0 &           &{\it11.50}  &         &K0V       	&210        &7.0 	&-21.9 & -14.7 & -6.5	&  0.6 &-139.0 & -36.4	&0.98 &T\\
HD 67945  	&08 09 38.6	&-20 13 50	&-37.8  &20.4 	&22~~~             &120~~~ &        &{\it 8.08}  &         &F0V       	&0          &14.3	&-22.1 & -13.4 & -4.2	&-34.7 &-60.0 & 8.5	&0.84 &T,SB2? \\
CD-58 2194	&08 39 11.6	&-58 34 28	&-32.7  &35.4 	&15.6              &85.0&{\it1.240 }&{\it10.18}  &{\it0.83}&G5V       	&270        & 9.8	&-21.9 & -16.7 & -5.4	& 8.3  &-100.3& -18.3	&0.98 &K  \\
CD-57 2315	&08 50 08.1	&-57 45 59	&-34.1  &31.0 	&11.7     	       &52.0&           &{\it10.21}  &         &K2Ve      	&308        & 9.3	&-22.1 & -13.0 & -5.5	& 9.2  &-105.8& -16.2	&0.90 &T  \\
TYC 8594-0058-1&09 02 03.9	&-58 08 50	&-24.2  &26.2 	&12.1              &34.1&           &11.30       &    0.83 &G8V       	&300        & 7.2	&-22.2 & -14.4 & -2.4	& 15.2 &-137.7& -18.7	&0.95 &   \\
CD-62 1197	&09 13 30.3	&-62 59 09	&-34.1  &27.4 	&12.7              &84.0&{\it0.54841}&10.46       &    0.93 &K0V(e)    	&280        & 8.4	&-21.7 & -16.3 & -5.8	& 22.1 &-114.9& -20.4	&0.91 &K  \\
TYC 7695-0335-1&09 28 54.1	&-41 01 19	&-28.7  &13.6 	&14~~~             &120~~~ &{\it0.3917}&{\it11.65}  &{\it1.10}&K3V     	&300        & 6.7	&-22.0 & -13.6 & -5.3	&-7.9  &-147.3&  18.8	&0.87 &K  \\
(*)BD-20 2977	&09 39 51.4	&-21 34 17	&-50.0  & 7.8 	&18.3              &10.1&           &{\it10.22}  &         &G9V       	&260        &11.3	&-22.3 & -16.5 & -4.7	&-21.9 &-79.0 &  34.4	&0.58 &T  \\
TYC 9217-0641-1&09 42 47.4	&-72 39 50	&-30.1  &18.9 	& 6.7              &23.4&{\it2.3026}&{\it12.30}  &{\it1.11}&K1V       	&240        & 6.5	&-22.1 & -13.6 & -6.2	& 50.4 &-139.6& -39.1	&0.88 &K  \\
CD-39 5833	&09 47 19.9	&-40 03 10	&-40.6  &16.3 	&15.0              &10.5&           &{\it10.89}  &         &K0V       	&260        & 9.0	&-22.2 & -15.7 & -4.6	&-2.1  &-109.5&  20.0	&0.92 &T  \\
HD 85151A 	&09 48 43.2	&-44 54 08	&-68.4  &33.5 	&14.3             	& &              &{\it 9.61}  &         &G7V       	&220        &15.3	&-22.3 & -15.8 & -3.9	& 2.5  &-64.8 &   7.7	&0.92 &T  \\
HD 85151B 	&09 48 43.4	&-44 54 09	&-68.4  &33.5 	&13.4             	& &              &{\it10.21}  &         &G9V       	&250        &15.3	&-22.3 & -14.9 & -4.0	& 2.5  &-64.8 &   7.7	&0.84 &T  \\
CD-65 817 	&09 49 09.0	&-65 40 21	&\textit{-32.3}  &\textit{19.9}& 7.3&19.2&{\it2.7388}&10.17       &0.75     &G5V       	&200        & 7.1	&-22.4 & -13.1 & -4.8	& 37.5 &-134.5& -22.5&0.97&T,K,D2.0"\\
HD 309851 	&09 55 58.3	&-67 21 22	&-42.4  &25.7 	& 6.6              &19.8&{\it1.8164}&9.90        &0.68     &G1V       	&170        & 9.2	&-22.5 & -13.2 & -4.5	& 31.8 &-102.6& -19.0	&0.98 &K  \\
HD 310316 	&10 49 56.1	&-69 51 22	&-43.5  &13.5 	& 4.7         	   &16.3&{\it3.625 }&10.07       &0.78     &G8V       	&224        & 8.2	&-22.4 & -13.6 & -5.2	& 46.5 &-110.7& -19.9&0.97&K,D0.6"\\
CP-69 1432	&10 53 51.5	&-70 02 16	&-32.6  &12.2 	& 3.6     	       &55.0 &{\it1.0305} &10.66     &0.71     &G2V       	&195        & 6.3	&-22.8 & -13.1 & -3.3	& 61.8 &-144.6& -26.1   &0.64 &K  \\
(*)CD-42 7422	&12 06 32.9	&-42 47 51	&-50.3  &-2.5 	& 3.3              &28.3 &{\it1.984 } &10.66     &0.92     &K0V       	&260        & 8.9	&-21.6 & -15.4 & -4.8	& 43.4 &-96.5 &  37.1   &0.50 &K  \\
CD-74 673   &12 20 34.4	&-75 39 29	&-109.4 & 4.2 	& {\it2.7}         &7.3        &{\it3.477}&{\it10.72}&{\it1.17}&K3Ve    &230        &19.5	&-21.7 & -15.5 & -2.8	& 25.8&-42.9&-11.5&0.75&K,3,4\\
CD-75 652 	&13 49 12.9	&-75 49 48	&-62.2  &-31.5	&-0.9              &20.1 &{\it2.2743} &9.67      &0.76     &G1V       	&200        &12.3	&-22.3 & -13.6 & -6.0	& 47.0 &-63.4 & -18.8   &0.93 &K  \\
HD 129496	&14 46 21.4	&-67 46 16	&-47.6  &-39.7	&-4.8             	 &88.0 &            &8.78      &0.60     &F7V       	&150        &11.1	&-22.5 & -13.6 & -5.9	& 61.6 &-64.9 & -11.5   &0.79 &   \\
NY Aps     &15 12 23.4	&-75 15 16	&{\it-73.9}  &{\it-73.1}	& 3.4    &10.8 &{\it4.084 } &{\it9.42} &{\it0.86}&G9V       	&182       &{\it19.9}&-21.2& -12.6 & -4.3	& 32.3 &-36.3 & -12.9 &0.62 &K,H\\
(*)HD 145689 	&16 17 05.4	&-67 56 29	&{\it-50.0}  &{\it-84.0}	&{\it-9~~~}&{\it106.4}&       &{\it5.95} &{\it0.18}&{\it A6V}    	&      &{\it19.2}&-22.7& -11.7 & -3.7	& 39.2&-32.4&-11.2&0.68&H,5,6,7 \\
CD-52 9381	&20 07 23.8	&-51 47 27	& 85.9  &-143.9&-13.3            	&42.0 &{\it0.8368} &10.59     &1.52     &K6Ve      	&60          &33.6	&-22.4 & -14.8 & -4.2	& 24.4 &-5.7  & -16.0   &0.80 &K  \\

\hline

\end{tabular}
}
\vfill
$^{1}$Data in italics are from the literature as follows: H~=~data from Hipparcos; T~=~data from TYCHO; K~=~data from \citet{Kiraga12}; 
(1) \citet{Riedel11}; 
(2) \citet{desidera}; 
(3) SB1, P$_{orb}$=613.9d \citet{Guenther07}; 
(4) \citet{covino97}; 
(5) Brown dwarf at 6.7" \citet{Huelamo10}; 
(6) \citet{zuckerman11}; 
(7) \citet{Diaz11}.\\ 
Entries marked (*) are new inclusions previously not listed by \citet{torres08}.

\end{table}
\end{landscape}

\begin{landscape}
\begin{table}
{
\caption{Revised list of high probability members proposed for the open cluster IC 2391$^{1}$ }
\label{table:argusic}
\begin{tabular}{lccrrrrlrclrrr@{\hspace{1.5pt}}r@{\hspace{1.5pt}}rr@{\hspace{1.5pt}}r@{\hspace{1.5pt}}rrl}


\hline

Name  &    $\alpha$&$\delta$&$\mu_\alpha$&$\mu_\delta$&$~~~V_r$&Vsini&P$_{phot}$&$V_J$ & B-V or&SpT &E$_{Li}$& $\pi_{kin}$&\multicolumn{3}{c}{$(U,V,W)$}&\multicolumn{3}{c}{$(X,Y,Z)$}&Prob&notes\\
     &        2000&2000 &\multicolumn{2}{c}{~~mas yr$^{-1}$}&\multicolumn{2}{c}{~km s$^{-1}$}&days&	 & $V-I_C$  &    &m\AA & mas&\multicolumn{3}{c}{km s$^{-1}$}&\multicolumn{3}{c}{pc}     &     &	 \\

\hline\hline
(*)PMM 5314 	&08 28 34.6	&-52 37 04	&-24.1&22.7 	&13.4     &63.0     &          &10.40    &    0.45 	&           &              &7.1    &-21.7 & -12.2 & -6.3   &-2.5  &-138.5& -19.6	&0.92 & \\
PMM 7422 	&08 28 45.6	&-52 05 27	&-22.6&24.6 	&14.8     &33.0     &{\it1.515}&10.49    &    0.69 	&G6             &233           &7.2    &-22.1 & -13.7 & -4.8   &-3.5  &-137.1& -18.6	&1.00&K\\
PMM 7956 	&08 29 51.9	&-51 40 40	&-23.0&18.8 	&15.3     &13.8     &          &10.62    &    0.98 	&e              &289           &6.5    &-21.3 & -13.8 & -7.7   &-4.6  &-152.7& -19.7	&0.71 & \\
PMM 6974 	&08 34 18.1	&-52 15 58	&-22.6&23.0 	&15.2     &5.2      &{\it7.80 }&12.26    &    1.04 	&               & 74           &6.9    &-22.1 & -14.4 & -5.1   &-2.1  &-143.5& -18.0	&1.00& 1 \\
PMM 4280 	&08 34 20.5	&-52 50 05	&-21.0&22.5 	&12.3     &16.0     &          &10.34    &    0.67 	&G5             &151           &6.6    &-22.1 & -11.7 & -4.3   &-0.9  &-151.0& -19.8	&0.90& \\
PMM 6978 	&08 35 01.2	&-52 14 01	&-23.3&22.3 	&15.2     &7.1      &{\it5.1  }&12.07    &    1.02 	&               &179           &6.9    &-22.0 & -14.3 & -5.7   &-2.0  &-143.6& -17.7	&1.00& 1 \\
PMM 2456 	&08 35 43.7	&-53 21 20	&-23.6&24.3 	&14.8     &48.0     &          &12.20    &    0.92 	&K3e            &301           &7.2    &-22.1 & -14.3 & -5.0   & 0.5  &-138.0& -18.4	&1.00&   \\
PMM 351  	&08 36 24.2	&-54 01 06	&-24.4&20.7 	&15.4     &90.0     &{\it1.923}&10.18    &    0.57 	&G0             & 90           &6.8    &-21.6 & -14.8 & -7.3   & 2.0  &-146.2& -20.3	&0.92& 1 \\
PMM 3359 	&08 36 55.0	&-53 08 34	&-23.1&23.6 	&14.7     &8.2      &{\it3.840}&11.51    &    0.76 	&               &226           &7.0    &-22.1 & -14.2 & -4.9   & 0.3  &-141.6& -18.2	&1.00& 1 \\
PMM 5376 	&08 37 02.3	&-52 46 59	&-20.8&20.9 	&15.6     &9.8      &          &14.30    &    1.37 	&e              &  0           &6.3    &-22.1 & -15.0 & -5.1   &-0.4  &-158.2& -19.7	&1.00&   \\
PMM 4324 	&08 37 47.0	&-52 52 12	&-22.4&22.0 	&14.3     &41.0     &          & 9.66    &    0.53 	&F5V            & 97           &6.7    &-22.1 & -13.8 & -5.2   & 0.0  &-148.9& -18.5	&0.97&   \\
PMM 665  	&08 37 51.6	&-53 45 46	&-21.1&22.7 	&14.5     &7.9      &{\it4.510}&11.35    &    0.75 	&G8e            &189           &6.5    &-22.2 & -14.3 & -4.4   & 1.9  &-151.7& -20.2	&1.00& 1 \\
PMM 4336 	&08 37 55.6	&-52 57 11	&\textit{-23.0}&\textit{24.9} 	&15.6 &8.0  &{\it0.371}&{\it11.55}&{\it0.98} &{\it G9}      &              &7.2    &-22.2 & -15.2 & -4.3   & 0.1  &-137.9& -17.2	&1.00&1,2\\
PMM 4362 	&08 38 22.9	&-52 56 48	&-23.5&22.4 	&15.4     &9.0      &{\it3.93 }&{\it10.95}&{\it0.74} &              &191           &6.9    &-22.0 & -14.7 & -5.6   & 0.2  &-144.1& -17.8	&1.00&1,2\\
PMM 4413 	&08 38 55.7	&-52 57 52	&-22.3&21.6 	&14.4     &8.6      &{\it5.05 }&10.31    &{\it0.74} &{\it 62}       &163           &6.6    &-22.1 & -13.9 & -5.2   & 0.4  &-150.9& -18.5	&1.00&1,2,3\\
PMM 686  	&08 39 22.6	&-53 55 06	&-21.3&21.7 	&15.2     &12.7     &{\it4.41 }&12.63    &    1.04 	&e              &215           &6.4    &-22.1 & -15.1 & -4.9   & 2.6  &-155.1& -20.4	&1.00&1  \\
PMM 4467 	&08 39 53.0	&-52 57 57	&-21.1&19.3 	&15.2     &12.6     &{\it3.85 }&{\it11.86}&{\it0.91}&{\it K0(e)}    &{\it190}      &6.1    &-21.9 & -14.7 & -5.9   & 0.7  &-164.0& -19.8	&1.00&1,2,4\\
PMM 1083 	&08 40 06.2	&-53 38 07	&-20.0&23.4 	&12.2     &67.9     &{\it1.333}&{\it10.45}&{\it0.69}&{\it G0}       &{\it162}      &6.5    &-22.2 & -12.2 & -2.9   & 2.1  &-152.6& -19.4	&0.92&1,2,4\\
PMM 8415 	&08 40 16.3	&-53 56 29	&-25.7&24.9 	&15.7     &20.7     &          &{\it11.84}&{\it0.94}&{\it G9(e)}    &{\it302}      &7.6    &-22.1 & -15.3 & -5.2   & 0.6  &-131.2& -15.7	&1.00&  2,4\\
PMM 1759	&08 40 18.3	&-53 30 29	&-26.2&21.2 	&13.5     &3.8      &          &{\it13.54}&{\it1.53}&{\it K4e}      &{\it55 }      &7.1    &-21.6 & -13.0 & -7.0   & 1.8  &-139.1& -17.4	&0.94&  2,4\\
PMM 1142	&08 40 49.1	&-53 37 45	&-18.4&23.6 	&14.0     &7.3      &          &{\it11.08}&{\it0.80}&{\it G1 }      &              &6.3    &-22.1 & -14.2 & -2.0   & 2.4  &-156.4& -19.6	&0.91&  2  \\
PMM 1174	&08 41 22.7	&-53 38 09	&-23.6&22.7 	&13.5     &60.0     &          &    9.54  &0.43     &F3V            & 79           &6.9    &-22.1 & -13.3 & -5.1   & 2.3  &-144.2& -18.0	&1.00&     \\
PMM 4636	&08 41 57.8	&-52 52 14	&-24.0&23.9 	&13.6     &4.9      &{\it5.80} &{\it13.57}&{\it1.55}&{\it K7e}      & {\it100}     &7.2    &-22.2 & -13.3 & -4.5   & 0.9  &-138.8& -16.0	&1.00& 1,2,4  \\
PMM 756 	&08 43 00.4	&-53 54 08	&-23.5&20.4 	&15.9     &16.5     & {\it3.14}&   11.16  &0.68     &G9             &217           &6.5    &-21.8 & -15.7 & -6.4   & 3.4  &-152.1& -18.9	&0.98&1    \\
PMM 5811	&08 43 17.9	&-52 36 11	&-21.3&21.6 	&14.3     &59.0     &          &    9.16  &0.37     &F2V            &107           &6.4    &-22.2 & -14.0 & -4.2   & 0.7  &-155.0& -17.0	&0.97&     \\
PMM 2888	&08 43 52.3	&-53 14 00	&-25.3&20.2 	&15.0     &66.5     &          &{\it9.76} &{\it0.56}   &{\it F5}    &              &6.9    &-21.6 & -14.6 & -7.0   & 2.1  &-144.7& -16.6	&0.96&2  \\
PPM 2012  &08 43 59.0 &-53 33 44  &\textit{-24.7}&\textit{25.1} &13.9&17.4   &{\it2.210}&{\it11.67}&{\it0.90}&{\it K0(e)}    &{\it251}      &7.4    &-22.2 & -13.9 & -4.3   & 2.6  &-134.5& -15.9  &1.00&1,2,4 \\ 
PMM 4809 	&08 44 05.2	&-52 53 17	&-21.3&19.7 	&13.8     &17.5     &{\it2.600}&{\it10.85}&{\it0.75}&{\it G3(e)}    &175           &6.1    &-22.1 & -13.5 & -5.2   & 1.6  &-162.6& -18.0	&1.00&1,2  \\
PMM 1373  &08 44 10.2 &-53 43 34  &\textit{-23.3}&\textit{24.4}&14.6&7.0     &{\it5.38} &12.25     &0.96     &               &156           &7.1    &-22.2 & -14.7 & -4.0   & 3.1  &-140.5& -16.8  &1.00&1    \\
PMM 5884 	&08 44 26.2	&-52 42 32	&-21.4&17.5 	&14.5     &13.9     &{\it3.03 }&{\it11.46}&{\it0.84}&{\it G9(e)}    &{\it225}      &5.8    &-21.8 & -14.1 & -6.5   & 1.4  &-170.2& -18.4	&0.98&1,2,4\\
PMM 4902 	&08 45 26.9	&-52 52 02	&-25.2&23.4 	&14.7     &7.7      &{\it4.820}&{\it12.76}&{\it1.24}&{\it K3e  }    &{\it204}      &7.2    &-22.1 & -14.5 & -5.1   & 1.6  &-137.5& -14.8	&1.00&1,2,4\\
PMM 6811 	&08 45 39.1	&-52 26 00	&\textit{-24.4}&\textit{24.8} 	&17.4 &90.0 &{\it0.653}&{\it 9.91}&{\it0.60}&{\it F8  )}    &137           &7.3    &-22.2 & -17.2 & -4.2   & 0.9  &-135.4& -13.9	&0.96& 2  \\
PMM 2182 	&08 45 48.0	&-53 25 51	&\textit{-22.4}&\textit{22.2} 	&14.7 &78.0 &{\it1.437}&10.22     &0.57     &               &187           &6.6    &-22.2 & -14.7 & -4.4   & 3.1  &-150.4& -17.0	&1.00&1    \\
\\

\hline

\end{tabular}
}
\vfill
$^{1}$Data mainly from \citet{platais07}; those in italics are from other literature sources as indicated (for proper motions, data in italics are from \citet{platais07}, otherwise from UCAC4; for color indices, data in italics represent V-I$_C$.):                   
 K~=~data from \citet{Kiraga12}; 
(1) \citet{messina}; 
(2) \citet{patten96};  
(3) SB2, P$_{orb}$=90.617d \citet{platais07}
(4) \citet{randich01}. \\
The entry marked (*) is a new inclusion previously not listed by \citet{torres08}.

\end{table}
\end{landscape}

\subsection{Kinematics and ages of IC 2391 and the Argus Association}

The Argus association  was easily discovered in the SACY survey due to its
special U velocity of $U=-22.0$km s$^{-1}$ as illustrated in Figure 2 of \cite{torres08}. 
The evidences presented in \cite{torres08}  strongly suggested that the Argus association and IC~2391 have a common origin. 
We have updated the SACY catalog in order to be able to test if members proposed by \cite{desidera}, 
\cite{zuckerman11}, and \cite{Riedel11} could be converged in to the IC2391 and Argus association member list.
Newly obtained data presented in this paper were also used.\\

The convergence was done using the new reduction of Hipparcos \citep{vanL}, the new proper motion catalog UCAC4 \citep{zacharias}
and the ASAS photometric survey \citep{Kiraga12}, for V and V-I colors, for stars not observed by us. 
We did not change the published proper motions for stars in the Hipparcos catalogue, and for double stars in UCAC4 with separations between about 1\arcsec\  to 10\arcsec\ because the image de-blending process in UCAC4 seems unreliable.
As we did in \citep{torres08}, we correct the photometric values for duplicity where possible. Note for the pair HD~85151A/B, there are no individual proper motions in Tycho-2. Therefore we used the primary proper motions for both stars. Our spectroscopic observations and the visual binary measurements from the Washington Double Star catalogue suggest they must be a physical pair.\\

The data for IC~2391 members are mainly from \cite{platais07} but using UCAC4 proper motions, except for double stars where we use \cite{platais07} and \cite{torres08} proper motions due to image de-blending issues with UCAC4 as explained above. There are 41 cluster stars with all kinematical data. Except for PMM~4413 whose systemic 
velocity is known, all other 20 known spectroscopic binaries were excluded from the study.\\

In Table~\ref{table:argus} and Table~\ref{table:argusic} we list the high probability members proposed for the Argus 
association stars and the open cluster IC~2391 respectively, obtained by applying the convergence method for both association and cluster stars together. 
We present a total of 65 high probability members which comprise of 32 Argus association stars and 33 IC~2391 members. We accept probabilities 
down to about 50 percent  as the stellar system is formed by the concentrated IC~2391 cluster and the 
widely scattered association stars around it, thus forming a spatial distribution that is far from gaussian. When we apply the convergence criteria on the association stars alone (without including IC~2391 members), then the probabilities of the association members increase significantly with the lowest membership probability being about 70 percent.)   \\

In order to test the true kinematical connection between IC~2391 and the Argus association stars, 
we also applied our convergence test to the two groups independently. 
The results are presented in Table~\ref{table:solution} and in Figure~\ref{fig:arxyz}. 1
The convergence method resulted in a near perfect match between the kinematics of IC~2391 stars and those of the Argus association stars from \citet{torres03}. The velocities of the two samples agree well within the errors. 
To give a better idea of the variations in the final $UVWXYZ$ values when 
the Argus association stars are considered together with the IC~2391 members, we have over-plotted the results obtained 
for the Argus field stars only (red squares), and the 
IC~2391 members only (blue triangles), with those obtained when the two groups are considered as one single association (gray circles). 
No significant difference is found.\\

\begin{table*}
\caption{Convergence method results for the Argus field stars and the members of open cluster IC~2391. }\label{table:solution}
\begin{tabular}{lrrrrrr}
\hline

Sample	& Distance [mas] &	U	[km/s] &	V	[km/s] &	W	[km/s] & N. members	&	Age [Myr] 	\\

\hline
Argus only (field) & 15.5 $\pm$ 19.9 &    -22.6 $\pm$ 0.4 &  -14.6 $\pm$ 1.5 &   -5.0 $\pm$ 1.2 &  32 &       26\\ 
IC~2391      	     & 6.9 $\pm$ 0.4~~~&  -21.7 $\pm$ 0.6 &  -14.2 $\pm$ 1.0  &  -5.1 $\pm$ 1.2 &  33 &       26\\
Argus \& IC~2391  & 11.2 $\pm$ 14.5    &-22.1 $\pm$ 0.4&   -14.4 $\pm$ 1.3&   -5.0 $\pm$ 1.2  &  65  &        26\\
\hline

\end{tabular}
\end{table*}

\begin{figure}
\begin{center}
\resizebox{\hsize}{!}{
\includegraphics[draft=False]{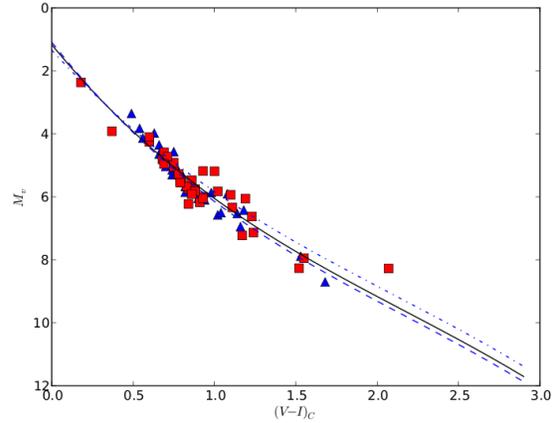}
 }
 \end{center}
\caption{The HR diagram of the members proposed for the Argus Association.
Blue triangles are the ArgA field members whereas the IC~2391 members are represented by red triangles.
The overplotted isochrones \citep{siess} are the ones for 20 (dot-dashed) , 30 (solid) and 40~Myr (dashed).}
\label{fig:arhr}
\end{figure}

Also as shown in Table \ref{table:solution} no significant difference is found in the ages of the two groups. 
Figure~\ref{fig:arhr} shows that the position of the Argus association in the HR diagram is consistent with a unique isochrone 
of 30~Myr $\pm$ 10~Myr, which is comparable to our obtained age of about 26~Myr. 
Our ages are from third degree polynomial isochorne curves \citep{torres08} based on some well observed star groups 
and compared partially with \cite{siess}. As they are not strictly calibrated, we must take these heuristic ages with caution, 
but of course the relative ages are more reliable. These results imply that both samples have the same age. 
Additionally, \cite{dasilva09} computed Li abundances for all SACY associations and showed that  the Lithium Depletion Pattern 
for the Argus association (defined by fitting the Li abundance as a function of effective temperature), fits well 
in a time sequence of the SACY associations with no difference between the Argus association stars and the IC~2391 members.


\subsection{Comments on {\bf excluded} stars}

Table~\ref{table:argus} and Table~\ref{table:argusic} are not the same as those presented in \cite{torres08}, 
as we used updated data and applied a new statistical model.  Below we comment on the excluded stars.\\

For the Argus association sample, 3 stars were excluded from the list of \cite{torres08}. Star BW~Phe was eliminated on kinematical grounds, star CD-49~1902 was eliminated for having too large value in Z and star HD~84075 has a U velocity well outside of the Argus association range. \\

For the IC~2391 sample, 3 stars were excluded from the list of \cite{torres08}. PMM~1560 and PMM~1820 were eliminated due to low membership probabilities based on kinematics. For PMM~3695  the proper motions are significantly distinct between \cite{platais07} and UCAC4, for unclear reasons. \\

\section{Elemental Abundances}\label{sect:3}

\subsection{High resolution observations}

In order to further explore the connection between the open cluster IC~2391 and the Argus association, we examine their elemental abundances. High resolution and high signal-to-noise spectra of members of the Argus association and IC~2391 were observed using 
VLT-UVES in the framework of program ID~082.C-0218 (PI Melo). Observations were undertaken in service 
mode using the UVES DIC\#1 (390+580nm) standard setting with a 0.8arcsec slit to achieve a spectral resolving power of R=60,000. 
The typical S/N ratio was $\sim$100 per pixel at 600~nm. The data were reduced with the latest UVES ESO-MIDAS pipeline. 
The resulting spectra were normalised using the $continuum$ task 
in the IRAF\footnote{IRAF is distributed by the National Optical Astronomy Observatory, which is operated 
by the Association of Universities for Research in Astronomy, Inc., under cooperative agreement with the National Science Foundation.} package. \\

We carry out a detailed elemental abundance analysis of six Argus association stars and eight IC~2391 stars with vsini $\leq$ 10 \kms. 
The other stars showed broadened spectral lines which meant there was greater blending between the lines, making individual line analysis more uncertain. Table \ref{t:results} lists the sample of stars analysed in this study together with their measured properties.

\subsection{Abundance analysis}\label{sec:analysis}
\begin{center}
\begin{figure*}
\includegraphics[width=12cm]{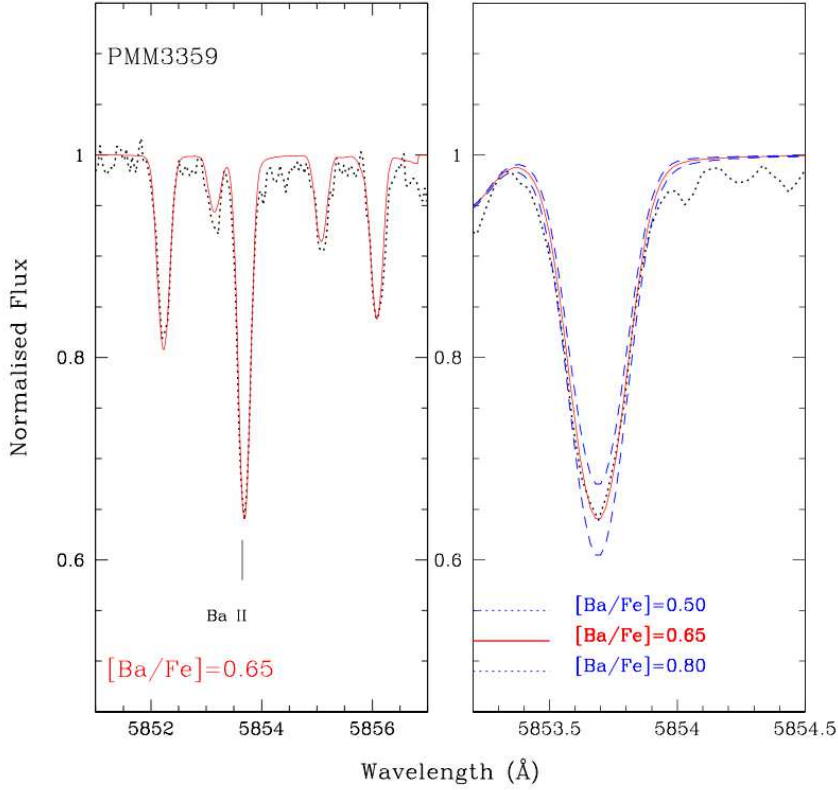}
\caption{Example of the Ba synthesis for star PMM~3359, with solid red line shows the best fit.}\label{f:basynt}
\end{figure*}
\end{center}

The elemental abundances were derived based on equivalent width (EW) measurements and spectral synthesis, 
making use of the latest version of the MOOG code \citep{sneden73}. 
The EWs were measured using the automated ARES code \citep{ares} with frequent manual checking of the EWs. 
Ba abundances were derived from the 5853~\AA\ line (which is known to be strong, isolated, un-blended and not affected by NLTE effects, 
e.g., \cite{mashonkina}) using the driver {\it synth} in MOOG. 
We retrieved hyperfine structure data from \citet{mw95}, adopting an isotopic solar mixture of 
81\% ($^{134}$Ba +$^{136}$Ba +$^{138}$Ba) and 19\% ($^{135}$Ba +$^{137}$Ba). 
An example of the synthesis of the Ba~{\sc ii} line at 5853 \AA\ is provided in Figure~\ref{f:basynt} for one of our sample stars (PMM~3359).\\

Interpolated Kurucz model atmospheres based on the ATLAS9 code \citep{Castelli97} with no convective overshooting were used throughout this study. 
We carried out the abundance analysis relative to the solar spectrum, taken with the same instrument and the same resolution of our sample stars. 
The full lines list, atomic data adopted and measured equivalent widths of a sample star (PMM~3359) can be found in Appendix~\ref{appendix:lines}. \\

We derive the stellar parameters based on spectroscopy. Abundances for all Fe {\sc i} and {\sc ii} lines were computed from the measured EWs. 
Effective temperature (\teff) was derived by requiring excitation equilibrium of the Fe~{\sc i} lines. 
Microturbulence was derived from the condition that abundances from Fe {\sc i} lines show no trend with EW. 
Surface gravity (\logg) was derived via ionisation equilibrium, i.e. requiring the abundances from Fe {\sc i} lines to equal those from Fe {\sc ii} lines. 
For stars cooler than 5500~K, we adopted a \logg\ value of 4.5, which is consistent with the cluster age. 
We did not attempt to optimize \logg\ via ionisation equilibrium for the cool stars because of the known over-ionisation 
effects (e.g., \cite{schuler2010}, see Section \ref{sec:res} for further discussion). 
The final adopted stellar parameters and the elemental abundances are given in Table \ref{t:results}. \\

\subsection{Error budget}

Chemical abundances are mainly affected by two kinds of uncertainties, namely $(i)$ errors due to the EW measurements 
or to the best-fit determination (in the case of spectral syntheses), and $(ii)$ errors due to the stellar parameters (T$_{\rm eff}$, log$g$, and $\xi$). 
The impact of the uncertainties on the atomic parameters, i.e. log $gf$, should be instead almost negligible since our analysis is
strictly differential with respect to the Sun (see Section~\ref{sec:analysis}).
Concerning the EW analysis, random errors in the [Fe/H] ratios due to EW measurements are well represented by the standard deviation (rms) 
from the mean abundance based on the whole set of lines, 
while the ones for the [X/Fe] ratios were computed by quadratically adding the rms for [Fe/H] and rms for [X/H]. 
Errors due to stellar parameters were estimated by varying one parameter at a time, and checking the corresponding variation 
in the resulting abundance. 
We adopted variations of $\pm$50~K in T$_{\rm eff}$ and $\pm$0.15 kms$^{-1}$ in $\xi$, because larger changes in those quantities 
would have introduced a significant trend of $\log~n$(Fe) $vs$ the excitation potentials and the line strength, respectively. 
For stars for which we could optimise the surface gravities, 
the uncertainties in $\log~g$ were estimated by varying this quantity until the difference between $\log~n$(Fe{\sc i}) 
and $\log~n$(Fe{\sc ii}) is larger than 0.1~dex, i.e. 
the ionisation equilibrium condition is no longer satisfied. 
The typical error in \logg\ was from 0.1 to 0.15. 
The final errors are then obtained by quadratically summing errors due to the EW measurements and the ones due to stellar parameters 
(see Table \ref{t:results}).\\

Focusing on abundances derived from spectral syntheses, we have errors due to the best-fit determination of typically 0.1~dex, 
reflecting uncertainties due to the continuum placement and to
the strong nature of the Ba features; similarly, since we are dealing with a very strong transition line, 
the errors due to stellar parameters are dominated by the microturbulence values ($\Delta$[Ba/Fe]=0.05-0.07~dex for $\xi$ changes of 0.15 \kms), while the effective temperatures and gravities have a minor impact, i.e. within 0.03~dex.

\subsection{Abundance results}
\label{sec:res}
\begin{center}
\begin{figure*}
\includegraphics[width=16cm]{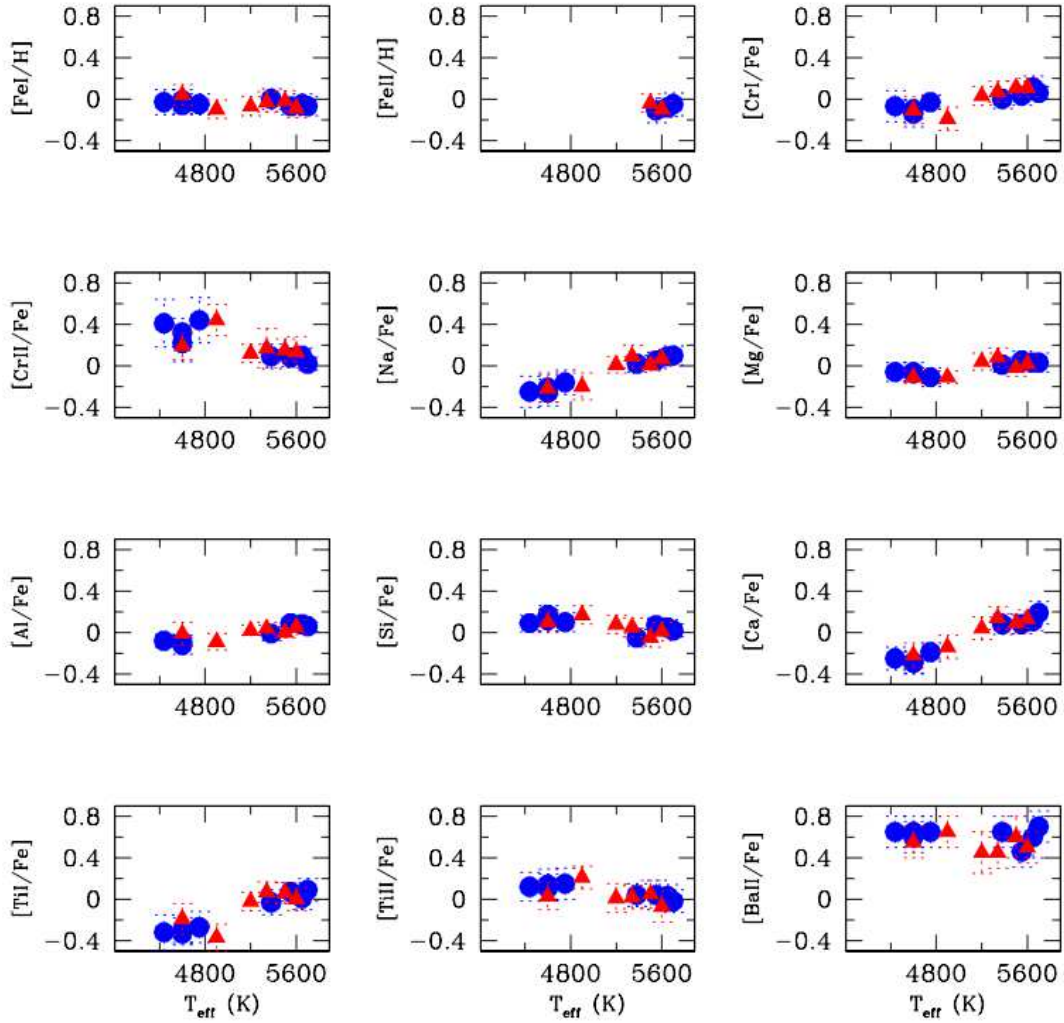}
\caption{[X/Fe] ratios as a function of the effective temperatures, showing the effects of over-ionisation. Note that [Ni/Fe] is not included as the Ni abundances do not show any temperature dependent trends (see Table~\ref{t:results}). Circles show IC~2391 stars and triangles show Argus association stars.}\label{f:ironpeak}
\end{figure*}
\end{center}

Final abundances are reported in Table~\ref{t:results}, where stellar parameters (T$_{\rm eff}$, log$g$, $\xi$) 
and [X/Fe] ratios along with the corresponding errors are provided for all our sample stars.
In Figure~\ref{f:ironpeak} we show the [X/Fe] ratios as a function of the effective temperatures 
for both IC~2391 (filled circles) and the Argus association (triangles).
The first evidence coming out from these plots is that 
we confirm previous findings by \cite{schuler03,schuler04,shen05,vdo09,schuler2010,biazzo2011}:  
over-ionisation/excitation effects are at work 
for stars cooler than T$_{\rm eff}$$\lesssim$5200 K. 
The photospheric abundances of Ca, Na, and Ti (from the Ti~{\sc i} lines)
show the most dramatic effect of this phenomenon, being 
about 0.2$-$0.3 dex lower in the cooler stars; similarly, Al and Cr (from the Cr~{\sc i} lines)
seem to show the same effect, although to a lesser extent.
Because we could also measure abundances for Cr and Ti from their first-ionisation stage transitions, 
we can confirm that we are observing an over-ionisation phenomenon: as can be seen in Table~\ref{t:results}, 
abundances from Ti~{\sc ii} and Cr~{\sc ii} lines result in larger values for the cooler stars, where the [X/Fe] ratios 
are enhanced up to 0.4~dex (see the [Cr~{\sc ii}/Fe] in the most extreme cases, e.g., PMM~4902 in IC~2391 or TYC~8561-0907-1 in Argus association). 
This fact reflects a pumping of the electrons 
to the ionised stage because of the strong UV flux coming from the hot chromospheres
(we refer the reader to \cite{schuler03, schuler2010, vdo09} for a more 
detailed discussion on this topic).
Si deserves special attention since we found  an over-excitation effect 
(all the employed lines have high excitation potentials, i.e. 4.92~$< \chi <$5.98~eV) acting on the cooler stars, 
that was also previously detected by \cite{schuler03} in M34, but not by \cite{vdo09}. 
Over-ionisation/excitation effects are also observed in the cool field stars (see e.g., \cite{allendeprieto}). Note that no temperature effects were observed for the Ni abundances in either the IC 2391 or the Argus association stars. Therefore the Ni abundance is not plotted in Figure~\ref{f:ironpeak}. \\

Due to the over-ionisation effects, the mean abundances and the associated rms error were computed considering all the stars for 
Fe~{\sc i}, Mg, Ni, and Ba, while we considered only the warmer stars (\teff $>$~5200K) when averaging [X/Fe] ratios for Na, Al, Si, Ca, Ti and Cr (see Table~\ref{t:results}). We found average metallicity of [Fe/H]=$-$0.04$\pm$0.03 and [Fe/H]=$-$0.05$\pm$0.04 for IC~2391 and Argus association, 
respectively: the cluster and the association exhibit a solar metallicity, with no evidence of internal scatter 
(the rms is significantly smaller than the observational uncertainties). 
Moreover, all the other $\alpha$ (Mg, Si, Ca, and Ti), odd-$Z$ (Na, Al), and iron-peak elements show solar ratios, 
pointing out that the cluster and the association are indistinguishable concerning their chemical composition 
(see Section~\ref{sec:disc} for a discussion on the scientific implication of our results).\\

Ba is notably over-abundant, with a mean of [Ba/Fe]=0.62$\pm$0.07 (IC~2391) and 
[Ba/Fe]=0.53$\pm$0.08 (Argus association). 
The slightly higher variation across the two stellar aggregates mainly reflect 
the larger uncertainties affecting Ba abundances. 
The Ba line feature is close to the saturation regime in the curve of growth, and as a consequence it is highly sensitive 
to the adopted microturbulence values. 
The extremely high Ba content that we obtained for our sample stars is not surprising and confirms
previous measurements (see Section~\ref{sec:disc}).
We stress, in passing, that over-ionisation effects can be ruled out as a possible explanation given that, 
at the effective temperatures of our sample stars, Ba is almost totally ionised. 
Further the possibility of NLTE effects is minimal as the 5853 line used in this study is not known 
to be affected by NLTE effects (\cite{mashonkina}).

\begin{landscape}
\begin{table}
\caption[]{Stellar parameters and abundances (see section \ref{sec:res} for details on the computation of the average values)}
\centering
\begin{tiny}
\begin{tabular}{lcccccccccccccccccr}
\hline\hline
star    & RV & vsini  & T$_{\rm eff}$ &  logg & $\xi$  &  [Fe~{\sc i}/H] &        [Fe~{\sc ii}/H]       &    [Na/Fe]     &     [Mg/Fe]    &     [Al/Fe]     &   [Si/Fe]     &     [Ca/Fe]     &      [Ti~{\sc i}/Fe]  &       [Ti~{\sc ii}/Fe] &     [Cr~{\sc i}/Fe]  &       [Cr~{\sc ii}/Fe] &       
[Ni/Fe]  &        [Ba~{\sc ii}/Fe]  \\
        &     \kms &  \kms &  (K)       &       & kms$^{-1}$    &                       &                &                &                 &               &                 &                &                 &                  &                 &                &                  \\
\hline
        &               &            &          &                       &                &                &                 &               &                 &                &                 &                  &                 &                &                   \\
        &           &               &       &            &          &                       &                &                &                  &      {\bf IC 2391}          &                 &               &                 &                  &                 &                &                   \\

        &               &            &          &                       &                &                &                 &               &                 &                &                 &                  &                 &                &                   \\
PMM 4902 &  14.7 & 7.10 &   4440 &   4.5 &   1.3  &	$-$0.03$\pm$0.12 &  .......  	& $-$0.25$\pm$0.15 & $-$0.06$\pm$0.09 &  $-$0.08$\pm$0.06 & 0.09$\pm$0.08 &  $-$0.25$\pm$0.11 & $-$0.32$\pm$0.17 &  0.12$\pm$0.14  &  $-$0.07$\pm$0.15  &	0.41$\pm$0.23 &  0.03$\pm$0.09 &  0.65$\pm$0.15\\
PMM 6974 &  15.1 & 4.99 &  4600 &   4.5 &   1.2  &	$-$0.06$\pm$0.08 &  .......  	& $-$0.21$\pm$0.14 & $-$0.06$\pm$0.09 &  $-$0.09$\pm$0.04 & 0.17$\pm$0.09 &  $-$0.29$\pm$0.10 & $-$0.33$\pm$0.09 &  0.14$\pm$0.15  &  $-$0.14$\pm$0.13  &	0.32$\pm$0.14 &  0.03$\pm$0.10 &  0.60$\pm$0.15\\
PMM 6978 &  15.4 & 6.77 &  4600 &   4.5 &   1.1  &	$-$0.01$\pm$0.10 &  .......  	& $-$0.26$\pm$0.13 & $-$0.07$\pm$0.05 &  $-$0.12$\pm$0.09 & 0.12$\pm$0.10 &  $-$0.29$\pm$0.11 & $-$0.31$\pm$0.13 &  0.11$\pm$0.13  &  $-$0.10$\pm$0.09  &	0.22$\pm$0.18 &  0.03$\pm$0.09 &  0.65$\pm$0.15\\
PMM 1373 &  13.9 & 6.68 &  4750 &   4.5 &   1.2  &	$-$0.05$\pm$0.10 &  .......  	& $-$0.16$\pm$0.12 & $-$0.11$\pm$0.09 &   .......	    & 0.10$\pm$0.09 &  $-$0.19$\pm$0.09 & $-$0.27$\pm$0.15 &  0.15$\pm$0.15  &  $-$0.03$\pm$0.07  &	0.44$\pm$0.22 &  0.02$\pm$0.10 &  0.65$\pm$0.15\\
PMM 3359 &  14.7 & 7.88 &  5380 &   4.5 &   1.2  &	   0.00$\pm$0.10 &  .......  	&  0.02$\pm$0.08 &  0.01$\pm$0.07 &  $-$0.01$\pm$0.07 &$-$0.05$\pm$0.08 &	0.08$\pm$0.10 & $-$0.03$\pm$0.12 &  0.04$\pm$0.11  &   0.00$\pm$0.09  &	0.09$\pm$0.12 & $-$0.03$\pm$0.11 &  0.65$\pm$0.15\\
PMM 665  &  14.6 & 7.47 &  5550 &   4.5 &   1.6  &	$-$0.07$\pm$0.08 & $-$0.11$\pm$0.12 &  0.05$\pm$0.08 &  0.05$\pm$0.09 &   0.09$\pm$0.04 & 0.07$\pm$0.07 &	0.08$\pm$0.09 &  0.07$\pm$0.09 &  0.04$\pm$0.14  &   0.03$\pm$0.09  &	0.08$\pm$0.09 &  0.03$\pm$0.09 &  0.46$\pm$0.15\\
PMM 4362 &  15.1 & 8.61 &  5650 &   4.3 &   1.4  &	$-$0.05$\pm$0.10 & $-$0.08$\pm$0.09 &  0.09$\pm$0.07 &  0.03$\pm$0.08 &   0.08$\pm$0.04 & 0.05$\pm$0.07 &	0.11$\pm$0.09 &  0.01$\pm$0.11 &  0.03$\pm$0.11  &   0.11$\pm$0.11  &	0.10$\pm$0.07 &  0.02$\pm$0.11 &  0.60$\pm$0.20\\
PMM 1142 &  14.0 & 6.71 &  5700 &   4.3 &   1.5  &	$-$0.07$\pm$0.09 & $-$0.05$\pm$0.11 &  0.10$\pm$0.09 &  0.03$\pm$0.09 &   0.06$\pm$0.04 & 0.02$\pm$0.06 &	0.08$\pm$0.09 &  0.09$\pm$0.11 & $-$0.02$\pm$0.11  &   0.06$\pm$0.06  &	0.02$\pm$0.06 & $-$0.01$\pm$0.10 &  0.70$\pm$0.15\\
        &         &	  &	   &			   &		    &		     &  	       &	       &		 &		  &		    &		       &		 &		  &		      \\
        &         &	  &	   &			   &		    &		     &  	       &	       &		 &		  &		    &		       &		 &		  &		      \\
{\bf AVERAGE $\pm$ rms}    &  &    &	  &	   &	$-$0.04$\pm$0.03   & $-$0.08$\pm$0.03& 0.06$\pm$0.04 & 0.03$\pm$0.02   & 0.06$\pm$0.05 & 0.02$\pm$0.05	 &	0.09$\pm$0.02 &	0.04$\pm$0.06 &	0.02$\pm$0.03&	0.05$\pm$0.05&  0.07$\pm$0.04 & 0.02$\pm$0.03	& 0.62$\pm$0.07	      \\

        &               &            &          &                       &                &                &                 &               &                 &                &                 &                  &                 &                &                   \\
        &       &               &           &            &          &                       &                &                &                 &      {\bf ARGUS}         &                 &                 &                 &                  &                 &                &                   \\
        &               &            &          &                       &                &                &                 &               &                 &                &                 &                  &                 &                &                   \\

CD$-$74 673   & 4.8 & 7.28 & 4600 &   4.5 &   1.3  & 0.04$\pm$0.10 &	.......		& $-$0.22$\pm$0.13 &  $-$0.11$\pm$0.06 &  $-$0.01$\pm$0.11 &  0.10$\pm$0.10 & $-$0.22$\pm$0.12 & $-$0.19$\pm$0.15 &   0.03$\pm$0.13 & $-$0.11$\pm$0.13 & 0.19$\pm$0.13 &  0.01$\pm$0.11 &	0.55$\pm$0.15 \\
TYC 8561-0907-1  & 15.6 & 4.41 & 4900 &   4.5 &   1.2  &$-$0.10$\pm$0.09 &	.......		& $-$0.20$\pm$0.13 &  $-$0.11$\pm$0.06 &  $-$0.09$\pm$0.08 &  0.17$\pm$0.09 & $-$0.14$\pm$0.11 & $-$0.37$\pm$0.13 &   0.21$\pm$0.11 & $-$0.19$\pm$0.11 & 0.44$\pm$0.15 &  0.01$\pm$0.09 &	0.65$\pm$0.15\\	
CD$-$42 2906  &23.5 & 10.35 &5200 &   4.5 &   1.6  &$-$0.07$\pm$0.09 &  .......	&  0.01$\pm$0.08 &   0.04$\pm$0.08 &   0.02$\pm$0.05 &  0.08$\pm$0.09 &  0.04$\pm$0.11 & $-$0.02$\pm$0.09 &   0.01$\pm$0.14 &  0.03$\pm$0.09 & 0.12$\pm$0.09 &  0.02$\pm$0.06 &	0.45$\pm$0.20\\
NY Aps      &  -3.2 & 10.95 & 5340 &   4.5 &   1.5  &$-$0.03$\pm$0.10 &	.......  &  0.09$\pm$0.11 &   0.08$\pm$0.09 &   0.04$\pm$0.06 &  0.05$\pm$0.09 &  0.14$\pm$0.11 &  0.07$\pm$0.10 &   0.02$\pm$0.11 &  0.07$\pm$0.10 & 0.17$\pm$0.19 & $-$0.03$\pm$0.07 &	0.45$\pm$0.15\\
CD$-$39 5833  & 14.9 & 10.17 & 5500 &   4.6 &   1.6  &$-$0.02$\pm$0.10 &  $-$0.04$\pm$0.09	&  0.01$\pm$0.06 &  $-$0.02$\pm$0.08 &   0.00$\pm$0.05 & $-$0.05$\pm$0.09 &  0.09$\pm$0.09 &  0.06$\pm$0.10 &   0.05$\pm$0.13 &  0.11$\pm$0.06 & 0.15$\pm$0.11 &  0.00$\pm$0.07 &	0.60$\pm$0.20\\
CD$-$28 3434  &  26.8 & 6.36 & 5600 &   4.3 &   1.5  &$-$0.09$\pm$0.09 &  $-$0.10$\pm$0.08	&  0.07$\pm$0.07 &   0.02$\pm$0.08 &   0.05$\pm$0.07 &  0.01$\pm$0.09 &  0.13$\pm$0.09 &  0.00$\pm$0.11 &  $-$0.07$\pm$0.15 &  0.11$\pm$0.09 & 0.13$\pm$0.15 &  0.00$\pm$0.06 &	0.50$\pm$0.15\\
        &         &	  &	   &			   &		    &		     &  	       &	       &		 &		  &		    &		       &		 &		  &		      \\
        &         &	  &	   &			   &		    &		     &  	       &	       &		 &		  &		    &		       &		 &		  &		      \\
{\bf AVERAGE $\pm$ rms}  &       &  &  &       &        & $-$0.05$\pm$0.05     & $-$0.07$\pm$0.04 & 0.05$\pm$0.04 & 0.03$\pm$0.04   & 0.03$\pm$0.02 & 0.02$\pm$0.06  & 0.09$\pm$0.05  & 0.03$\pm$0.04 & 0.00$\pm$0.05 & 0.08$\pm$0.04 & 0.14$\pm$0.02 & 0.00$\pm$0.02 & 0.53$\pm$0.08\\
\hline
\hline
\end{tabular}
\end{tiny}
\label{t:results}
\end{table}
\end{landscape}

\subsection{Comparison with previous abundance studies}
The chemical composition of IC~2391 has been previously determined by two different studies, 
namely Platais et al. 2007 (P07\nocite{platais07}) and \cite{vdo09}. \\

In P07, the authors presented iron abundances for a sample of four G-type members, obtaining an average metallicity 
of [Fe/H]=0.05$\pm$0.05, which is in very good agreement with our estimate. 
Three out of four P07 stars are in common with our study, i.e., PMM~3359, PMM~665, and PMM~4362; 
the differences in effective temperatures are $\Delta$T$_{\rm eff}$(our$-$p07)=100~K, 99~K, and 34~K, respectively, 
while gravities agree within 0.05 dex. Microturbulence values agree very well for the star PMM~3359 ($\Delta\xi$=0.03 km s$^{-1}$) 
and PMM~4362 ($\Delta\xi$=0.1 km s$^{-1}$), while P07 derived a slightly lower value for the star PMM~665,
i.e., 1.25~km s$^{-1}$ to be compared with our estimate of 1.60~km s$^{-1}$.\\

\cite{vdo09} presented abundances of Na, $\alpha-$ (Si, Ca, and Ti), and iron-peak elements 
(Fe, and Ni) for a sample of seven IC~2391 G/K-type members. 
They found the following average values: [Fe/H]=$-$0.01$\pm$0.02, [Na/Fe]=$-$0.02$\pm$0.03, [Si/Fe]=0.01$\pm$0.02, 
[Ca/Fe]=0.02$\pm$0.01, [Ti~{\sc i}/Fe]=0.00$\pm$0.02, [Ti~{\sc ii}/Fe]=0.14$\pm$0.13, and [Ni/Fe]=0.00$\pm$0.02.
Comparing their results with our mean abundances (see Table~\ref{t:results}) we can conclude 
that the two studies are in excellent agreement; the only exception 
is the Ti abundances derived from the Ti~{\sc ii} lines, for which they obtained a 0.12~dex higher value. 
Note that their [Ti~{\sc ii}/Fe] average was based on all sample stars, including those with over-ionisation effects. 
Should we include all sample stars when determining the average, then our abundance estimates would agree well.
Two of our sample stars, PMM~4902 (VXR~76a) and PMM~4362 (VXR~3), were already analysed by \cite{vdo09}. 
In Table~{\ref{t:compdr09} we compare our results with the ones by \cite{vdo09}:
the two determinations agree within their respective uncertainties for all stellar parameters and final abundances. We note that 
the higher discrepancy for the Ti abundance from the Ti~{\sc ii} lines is still within the uncertainties, 
which are significantly larger in this case (a different set of line list could partially explain such a difference).
The barium abundance for IC~2391 is presented in \cite{vdoba}, where they derived a mean abundance
of [Ba/Fe]=0.68$\pm$0.07, which is in excellent agreement with our value ([Ba/Fe]=0.62$\pm$0.07). \\

The chemical content of the Argus association was previously determined by \cite{viana}, who gathered Fe, Si and Ni abundances 
for a sample of seven members. 
They also conclude that the association shows a solar abundance pattern, finding 
[Fe/H]=$-$0.03$\pm$0.05\footnote{We quote their uncorrected value, and refer the reader to that paper 
for details on the trends with effective temperatures among their sample stars.}, [Si/Fe]=$-$0.03$\pm$0.03, and [Ni/Fe]=0.02$\pm$0.04. 
We have four stars in common with the work by \cite{viana}, namely CD-28~3434, CD-42~2906, TYC~8561-0970-1, and CD-38~583. 
Differences in T$_{\rm eff}$ are within 100K for all the stars with the exception of TYC~8561-0970-1 for which we derived a T$_{\rm eff}$=4900~K, 
while their final estimate is T$_{\rm eff}$=5348~K. 
However, our final value agrees better with the spectral type K0 found by \cite{torres06} 
(see Table~\ref{table:argus}) compared to that by \cite{viana}. 
Moreover, while gravity values agree well (with an average difference of 0.125), larger discrepancies are seen for the microturbulence values. 
\cite{viana} derived systematically higher microturbulences (on average about 0.15~kms$^{-1}$ higher, 
which is still in fair agreement within the uncertainties) and noted that their higher $\xi$ values are probably due to 
a different set of spectral lines, which might be particularly 
sensitive to the strong magnetic field characterising these stars, resulting in larger $\xi$ values. 
Whether young pre-main sequence stars exhibit higher microturbulence values is still under discussion 
and there is no general consensus on this topic (\cite{steenbock, james06}). 
Finally, [Fe/H], [Si/Fe] and [Ni/Fe] agree well among the two studies, with differences always smaller than 0.1~dex. \\

\begin{center}
\begin{table*}
\caption{Comparison between our study and the one by D'Orazi \& Randich (2009) for two stars in common.}\label{t:compdr09}
\begin{tabular}{lcccccccccr}
\hline\hline
star                    & Teff & logg & $\xi$ & [Fe/H] & [Na/Fe] & [Si/Fe] & [Ca/Fe] & [TiI/Fe] & [TiII/Fe] & [Ni/Fe] \\
                        &  (K) &      &  (kms$^{-1}$) & &         &        &          &          &          &          \\
\hline
                        &      &	&	      & &         &        &          &          &          &       \\
PMM4362                 & 5650$\pm$50 & 4.30$\pm$0.10  & 1.40$\pm$0.15 & $-$0.05$\pm$0.10 & 0.09$\pm$0.07 & 0.05$\pm$0.07 & 0.11$\pm$0.09 &
0.01$\pm$0.11 & 0.03$\pm$0.11 & 0.02$\pm$0.11\\
PMM4362$_{\rm DR09}$   &  5590$\pm$60 & 4.45$\pm$0.10 & 1.15$\pm$0.15 & 0.00$\pm$0.08 & $-$0.02$\pm$0.07 & 0.02$\pm$0.07 & 0.01$\pm$0.05 &
0.00$\pm$0.07 & 0.03$\pm$0.10 & $-$0.02$\pm$0.07 \\
  & & & & & & & & & & \\
\hline
 & & & & & & & & & & \\
PMM4902                & 4440$\pm$50 & 4.50$\pm$0.15 & 1.30$\pm$0.15 & $-$0.03$\pm$0.12 & $-$0.25$\pm$0.15 & 0.09$\pm$0.08 & $-$0.25$\pm$0.11 &
$-$0.32$\pm$0.17 & 0.12$\pm$0.14 & 0.03$\pm$0.09\\
PMM4902$_{\rm DR09}$  & 4343$\pm$60 & 4.50$\pm$0.10 & 1.20$\pm$0.15 & 0.00$\pm$0.09 & $-$0.34$\pm$0.11 & 0.06$\pm$0.12 & $-$0.38$\pm$0.14 &
$-$0.35$\pm$0.17 & 0.37$\pm$0.14 & 0.02$\pm$0.12\\
\hline\hline
\end{tabular}
\end{table*}
\end{center}

\section{Discussion}\label{sec:disc}

\begin{center}
\begin{figure*}
\includegraphics[width=12cm]{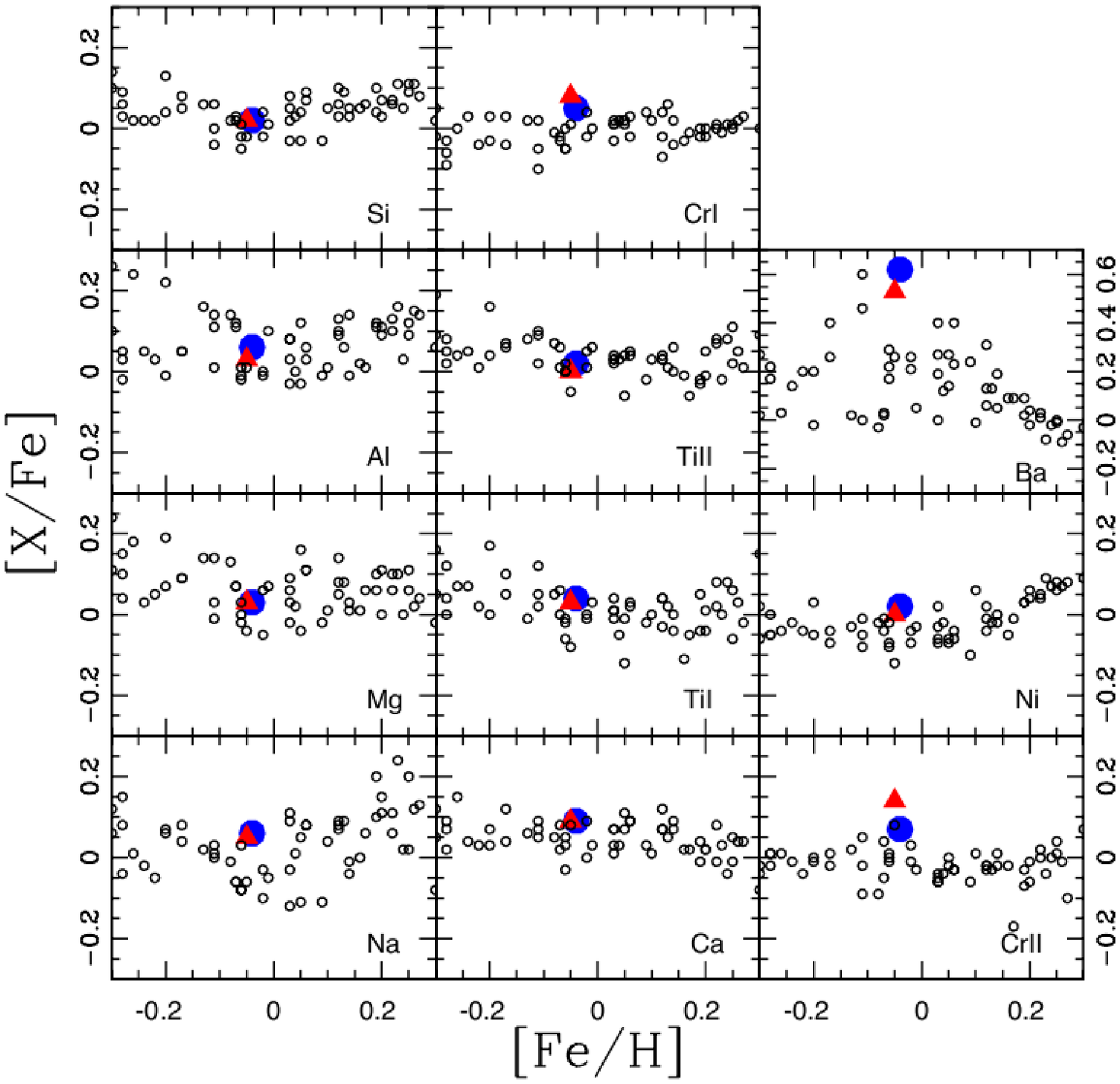}
\caption{The average abundances for IC~2391 cluster (blue circle) and Argus association stars (red triangle) compared to 
the solar neighborhood abundances of Bensby et al. (2005) (open circles). The average abundance values plotted here do not include the stars which show effects of over-ionisation. Refer to section \ref{sec:res} for details on how the average abundances were derived}
\label{f:fieldgrid}
\end{figure*}
\end{center}

\subsection{Chemical tagging}

Stars born within a single star-forming aggregate share its chemical signature with all stars in the cluster, 
where the elemental abundance pattern represents the conditions of the proto-cluster gas cloud. 
High resolution studies targeting a large range of chemical elements have confirmed this to be true among several Galactic open clusters \citep[e.g][]{gds06, cr261, pancino2010}. 
Indeed the results presented in Table \ref{t:results} show that the open cluster IC~2391 is also highly chemically homogeneous, 
with any abundance scatter being well within the measurement uncertainty.\\

The technique of chemical tagging proposes to use the elemental abundance patterns of stars to identify dispersed field stars 
with their original formation site \citep{fbh}. 
Several recent examples show the concept of chemical tagging at work; the HR~1614 moving group \citep{hr1614}, 
the Hercules stream \citep{hercules}, the Wolf~630 moving group \citep{wolf630}, and the Hyades Supercluster \citep{hysc, pompeia}. 
In all these examples the dispersed stellar system was identified via kinematical information and its reality as a disrupting cluster 
was confirmed (or disproved in the case of the Hercules stream) via chemical abundances. 
Similarly, the presence of the Argus association was identified by the stellar kinematics in the SACY survey 
and has been kinematically associated with the open cluster IC~2391, see Figure \ref{fig:arxyz}. 
Probing the chemical information of their stars, we show in Table \ref{t:results}, the average elemental abundances 
between the open cluster stars  and the dispersed members of the Argus association, are almost identical within the uncertainties. 
This similarity holds for all studied elements, 
which include those formed via the $\alpha$-process, Fe-peak elements formed via nuclear fusion and $s$-process elements 
formed via neutron capture. 
This seems to support the case that the stars in the Argus association originally formed together with the open cluster IC~2391. 
Note that unlike the previous cases of 
chemically tagged stellar structures, IC~2391 and the Argus association are young stellar systems, with an age of $\sim$30~Myr as presented in Table~\ref{table:solution}).\\

We now compare the cluster and association abundances against the solar neighborhood abundances published by \cite{bensby05}. 
As seen in Figure \ref{f:fieldgrid}, most of the abundances of the studied structures sit well within the solar neighborhood values. 
Therefore this lessens the strength of the chemical tagging argument discussed above, 
as other field stars may also share a similar chemical abundance pattern as that of the IC~2391. 
Applying the method of chemical tagging we consider the number of stars in the \cite{bensby05} sample which share 
the same abundance pattern as those of IC~2391, with the exception of Ba. Keeping a margin of $\pm$0.1~dex 
we find that 10 stars out of the 102 in the \cite{bensby05} sample could be classed as sharing the same abundance patterns (except Ba). 
If we include Ba abundances in the chemical tagging process, then only 1 star in the \cite{bensby05} sample share a common chemistry as IC~2391. 
Should we adopt a more strict criteria with an abundance margin of $\pm$0.05~dex, which is the level of homogeneity seen within bound clusters, 
then none of the \cite{bensby05} stars share the open cluster abundance pattern, even without considering Ba. 
Clearly abundance matching of this form is plagued by systematic effects between different studies. 
We await the availability of data from large surveys such as the GALAH survey \footnote{www.aao.gov.au/HERMES/GALAH} planned with the HERMES instrument \citep{barden},
which would be homogeneously observed and analysed, to further expand this field of research.\\

\begin{center}
\begin{figure*}
\includegraphics[width=12cm]{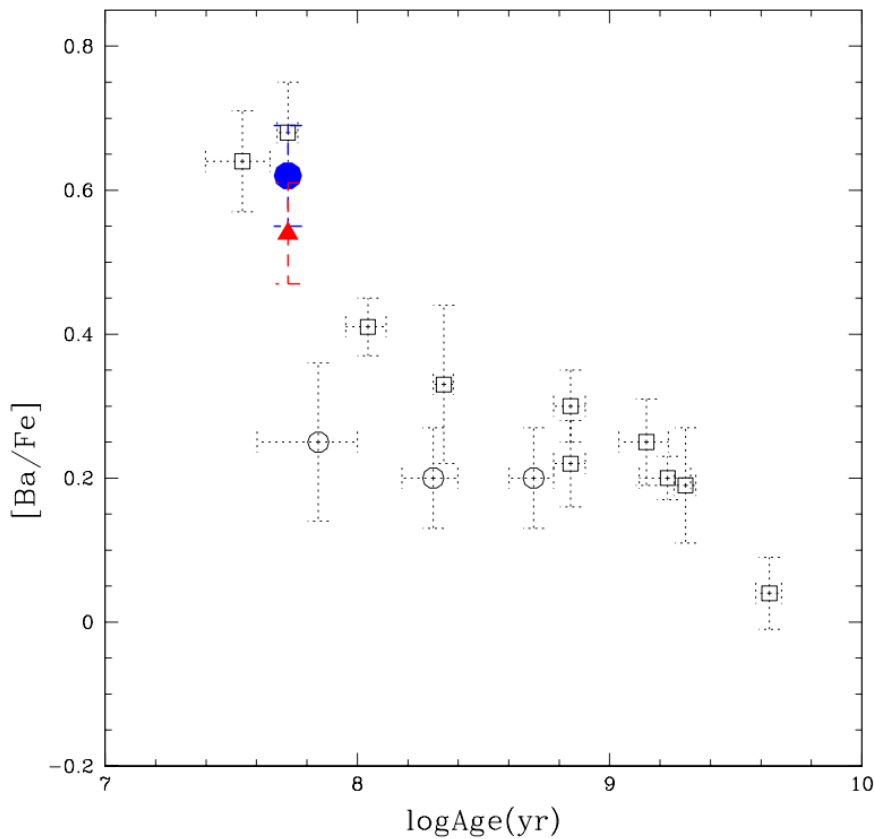}
\caption{[Ba/Fe] as a function of cluster age for {\bf IC~2391 (filled circle) }and the Argus association (filled triangle) 
as well as open clusters (labelled as open squares from D'Orazi et al. 2010) and moving groups (labelled as open circles from D'Orazi et al. 2012).}
\label{f:BaAge}
\end{figure*}
\end{center}

\subsection{Ba abundances}

The abundance of Ba, the only analysed $s$-process element in our study, is clearly over-abundant with respect to the field as noted in the discussion above. Similar enhancement in Ba is seen in other young clusters, 
including the Hyades cluster and supercluster (\cite{gds06, hysc}). \cite{vdoba}
detected, in a sample of 20 open clusters, an anti-correlation between the [Ba/Fe] ratios and the cluster age.
While OCs with ages $\gtrsim$4~Gyr are characterised by solar Ba abundances, the OCs with ages $\sim$100-200~Myr 
exhibit an enhancement up to 0.2-0.3~dex. 
This increasing trend can be reproduced only if the contribution from low-mass AGBs (1.0$<$M/M$_\odot$$<$1.5) 
to the Galactic chemical enrichment is higher than what was previously thought (see \cite{vdoba,maiorca} for further details).
Most intriguingly, the younger clusters of their sample ($\lesssim$70~Myr) show an even higher Ba content
([Ba/Fe] up to 0.5-0.6~dex), which cannot be explained {by} this scenario. 
It is quite unlikely {\bf that} an enrichment in the Ba content took place in the last $\sim$100~Myr of Galactic evolution, 
unless mechanisms of local enrichment are invoked. Similarly, 
\cite{desidera}, analysing the star HD~61005 and its possible link to the Argus association, 
confirm the same abundance pattern, i.e. the [Ba/Fe] is more than a factor of four above the solar value. 
In Figure \ref{f:BaAge} we report the run of [Ba/Fe] with age for clusters presented 
in \cite{vdoba}\footnote{Due to a systematic offset between dwarfs and giants in the \cite{vdoba} sample, 
here we consider, for homogeneity, only clusters whose 
abundances come from un-evolved members.}, for the three young moving groups by \cite{vdo12}, 
along with our measurements for IC~2391 and Argus field stars (this study). 
As one can see, our values fit the Ba-age relationship very well, confirming that very young open clusters seem to share 
an extremely high Ba abundance.\\

Moreover, the peculiar nature of Ba stands out from the other $s$-process elements: when available, 
abundances of Y, Zr (first-peak $s$-process element), and La,Ce
(second-peak $s$-process element) show solar ratios (see
\cite{gds06,carrera} for the Hyades; \cite{vdo12} for three moving groups AB~Doradus, Carina-Near and Ursa-Major).
Since none of the current available models can reproduce such a trend (accounting for an over-production of Ba without a similar 
trend in La and/or Ce) we are probably dealing with 
several spurious, conspiring effects which mimic a super-solar Ba content. 
A wide discussion of this topic is out of the purpose of the present paper 
(we refer the reader to \cite{vdo12}), here we just recall that the presence of hot chromospheres in these young stars might play, 
directly or indirectly, a certain role in the 
strength of the Ba~{\sc ii} line(s).\\

\subsection{The dissolution of IC 2391}
With all present evidence pointing to a common origin of IC 2391 and the Argus association stars, we now explore the likely dissolution mechanism of the original open cluster. \cite{lamers06} have shown that most of the energy necessary to dissolve a cluster comes from collisions with giant molecular clouds. Depending on the impact parameter and the energy involved in the collision, a halo of stars is left around a central spheric or more oblate core. Qualitatively speaking this is what we see in Figure~\ref{fig:arxyz} with the IC 2391 being the core and the Argus association stars as part of the halo \citep{gieles06}. The lack of symmetry around the cluster (blue triangles), however, is partially due to the fact that the IC 2391 is on the limit of the SACY survey \citep{torres08} of 150~pc. The second half of the halo being beyond 150~pc.\\

For a small-N system like IC~2391 a halo of unbound stars can also be the result of ejections of stars from the cluster core due to two-body encounters. Even when the tidal field is weak, clusters lose stars because of close encounters of stars in the cluster. 
This process is important for clusters whose age exceeds the half-mass relaxation time $\trh$.
Based on an estimated half-mass radius of about $1.5$~pc \citep{2002A&A...389..871D}, 
a mass of about 500~$\msun$ \citep{2009ApJ...706.1484B} we use the standard expression for the half-mass relaxation time-scale 
$\tau_{\rm rh}$ of \citet{1971ApJ...164..399S} and find $\trh\simeq 70$~Myr. 
Here we assumed the mean mass of the stars to be $0.5\,M_\odot$ (i.e. $N=1000$) and we use $0.02N$ as the argument of the Coulomb logarithm. 
Although this value for $\trh$ is uncertain to within a factor of two or three, the similarity between $\trh$ 
and the age of IC~2391(about  30$\pm$10~Myr) suggests that it is possible that IC~2391 is dynamically evolved 
and has ejected a fraction of its stars. These unbound stars we now observe as the Argus association around the cluster. 
It implies that the cluster was  smaller and denser in the past \citep{1965AnAp...28...62H, 2010MNRAS.408L..16G}. \\

In fact, we can speculatively estimate the initial size and velocity dispersion of the cluster. 
The radius of the Argus associations is about 150~pc. 
Combined with an age of 30~Myr for IC~2391  this means that the most distant stars must have escaped with a velocity of about 5~\kms. 
The typical velocity with which stars escape from an isolated cluster is few $\vrms$, where $\vrms$ is the root-mean square velocity of cluster.  
So the most distant stars escaped when the cluster had $\vrms\simeq$1-2~\kms. 
Combined with the mass this  corresponds to  a half-mass radius of about $\rh$$\simeq$0.2-0.4~pc. 
In Fig.~\ref{fig:nbody} we illustrate this idea with the result of an $N$-body simulation of an isolated cluster with an initial $\rh$=0.25~pc. 
After 30~Myr there are unbound stars out to 100~pc away from the cluster, which have velocities of a several \kms. 
The current $\rh\simeq1.1\,$pc and 897 stars are still bound to the cluster, comparable to IC~2391.\\

This implies that the initial density of IC~2391 was several orders of magnitude higher than the present day density 
and the initial relaxation time a factor of a few shorter. 
This could be the explanation for the observed mass segregation in the cluster \citep{2009ApJ...706.1484B}. 
These are of course very rough estimates, because we have ignored the presence of the Galactic tidal field 
and external factors such as giant molecular clouds.

\begin{center}
\begin{figure*}
\includegraphics[width=12cm]{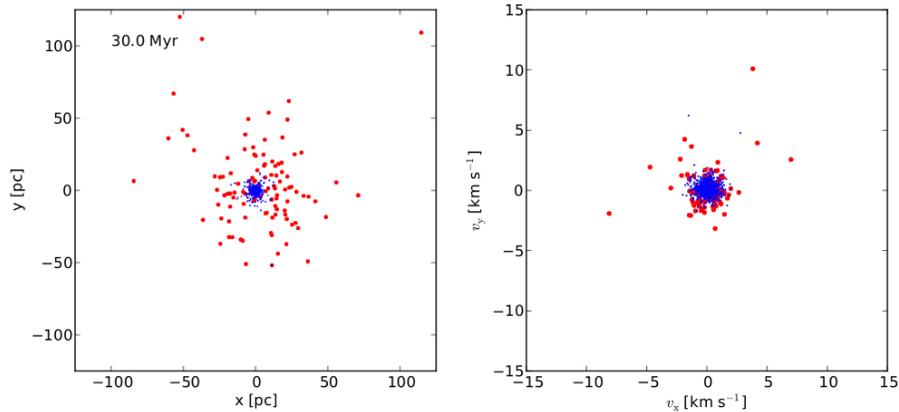}
\caption{Results of an N-body simulation with NBODY6 
(Aarseth 2003) of an isolated cluster with N=1024, a Kroupa (2011) stellar mass function between 
0.1 $M_\odot$ and 10 $M_\odot$. The initial density profile followed from a Plummer (1911)
model with a half-mass radius of $r_{\rm h} = 0.25$ pc. 
Small blue symbols represent all bound stars at an age of 30 Myr and the larger red circles are unbound members. The cluster and halo are comparable to IC 2391 and the Argus association at 30~Myr.}\label{fig:nbody}
\end{figure*}
\end{center}

\section{Summary and conclusions}
In this paper, we presented the members of the Argus association as identified via the SACY survey and find the age of the system 
to be about  30$\pm$10~Myr.  
Based on high resolution UVES spectra we determined the abundances for Fe, Na, Mg, Al, Si, Ca, Ti, Cr, Ni and Ba 
in both the open cluster IC~2391 and the Argus association. 
All stars in the open cluster and Argus association were found to share similar abundances with the scatter well within the uncertainties, 
where [Fe/H] = $-$0.04 $\pm$ 0.03 for cluster stars and 
[Fe/H] = $-$0.06 $\pm$ 0.05 for Argus stars. All other elements were in their solar proportions 
with the exception of Ba which we observed to be enhanced to around 0.6~dex. We also discussed the effects of 
over-ionisation/excitation observed in our sample for stars cooler than about 5200~K, as has been previously noted in the literature. \\

In summary we present strong kinematic, evolutionary and chemical evidence to support a 
common origin for the Argus association and the open cluster IC~2391. In particular 
the chemical tagging of the stars presented in this paper provides {\em a necessary} condition 
to prove the link between the Argus association and IC~2391. However, as discussed in the text, 
many elemental abundances of the studied stars are indistinguishable from the solar neighborhood, 
with the exception of Ba which is an important piece of evidence in favor of a common origin of 
the two stellar systems. For reasons which are not yet completely understood, barium abundances have been recently 
found to correlate with age. The abundances found for both stellar systems are similar. Moreover, they fit the Ba-age correlation very well. \\
 
Given all evidence in hand, we conclude that we are witnessing the on-going dissolution of IC~2391, 
where the stellar members of the Argus association stars were originally born in the same proto-cluster cloud as IC~2391. 
Simple modeling of this system finds the dissolution of this timescale to be consistent with two-body interactions. 
A more detailed modeling of the disruption mechanisms, predicting spatial distribution and possibly the velocity 
dispersion of both groups would strengthen the proposed scenario. From an observational point of view, 
the second half of the stellar halo will be easily found by GAIA settling the question.

\section*{Acknowledgments}
C. Melo would like to thank the AAO Distinguished Visitor Program and the ESO DGDF for funding his visit to AAO/Australia. 
C. Torres and C. Melo are grateful to the ESO DGDF and Visitor Program for continuous support for to the SACY project 
in the form of DGDF, science-leaves and visitorships. MG thanks the Royal Society for financial support. We thank the anonymous referee for helpful suggestions and Simon O'Toole for proof reading the manuscript.  \\

We thank the Centre de Donn\'{e}es Astronomiques de Strasbourg (CDS), the
U. S. Naval Observatory and NASA for the use of their electronic facilities,
especially SIMBAD, UCAC4, the Washington Double Star Catalog (WDS) and ADS.


\appendix

\section[]{Atomic line list}\label{appendix:lines}

\begin{center}
\begin{table}
\caption{Atomic line list. Full table data available online.}
\begin{tabular}{lcccccccccr}
\hline\hline
Element & Wavelength (\AA) & LEP (eV) & log $gf$ & EW PMM3359 (m\AA)  \\
\hline
Na\,{\sc i}	&	5682.63	&	2.1	&	-0.71	&	123.2	\\
Na\,{\sc i}	&	5688.21	&	2.1	&	-0.40	&	144.7	\\
Na\,{\sc i}	&	6154.23	&	2.1	&	-1.57	&	56.0	\\
Mg\,{\sc i}	&	5711.09	&	4.35	&	-1.71	&	126.5	\\
-- &  -- & -- & -- & --\\
-- &  -- & -- & -- & --\\
-- &  -- & -- & -- & --\\
\hline\hline
\end{tabular}
\end{table}
\end{center}

\end{document}